\documentclass[onecolumn,draftcls]{IEEEtran}

\usepackage{mleftright}       
\mleftright                   

\usepackage{amsmath,booktabs,bm,xcolor}         

\usepackage{cite} 

\IEEEoverridecommandlockouts

 \title{Compression of Sources with Memory and Private Random Access}

 \author{%
 Venkat Chandar\thanks{V.\ Chandar is with DE Shaw, New York, USA. A.\ Tchamkerten is with the Institut Polytechnique de Paris, Palaiseau, France. S.\ Vatedka is with the Indian Institute of Technology Hyderabad, Sangareddy, Telangana, India.},
 Aslan Tchamkerten, Shashank Vatedka
    \thanks{The work of Shashank Vatedka was supported by the project ANRF/ARGM/2025/000928/TS from ANRF, India.}
    \thanks{All mathematical arguments and proofs in this paper were
developed without AI assistance. Any errors are therefore
entirely human.}
 }

\newcommand{\svcomment}[1]{\textcolor{red}{SV: #1}}
\newcommand{\atcomment}[1]{\textcolor{red}{AT: #1}}

\usepackage{basicreq,bbm,xcolor}

\newcommand{\openone}{\leavevmode\hbox{\small1\normalsize\kern-.33em1}}

\newcommand{\ie}{{\textit{i.e.}}}

\newcommand{\bo}[1]{{\bm{#1}}}

\newcommand{\supp}{\mathrm{Supp}}

\IEEEaftertitletext{\vspace{-1.5\baselineskip}}
\let\underbrace\LaTeXunderbrace

\begin{document}
\maketitle

\begin{abstract}
Private local decoding imposes a stringent constraint on compression: the codeword bits used to recover a source symbol must reveal nothing about the remaining symbols beyond what the recovered symbol itself reveals. For memoryless sources, this constraint is known to be compatible with entropy-rate compression. We show that the same holds for any stationary ergodic source over a finite alphabet. Specifically, every rate above the entropy rate is achievable with exact privacy and error-free local decoding on a set of source sequences whose probability tends to one. In contrast, we exhibit non-ergodic sources for which private local decoding incurs a strictly positive rate penalty. 
\end{abstract}

\section{Introduction}

Let $\{X_i\}_{i\geq 1}$ be a stationary ergodic source
over a finite alphabet, and let
\[
\bo X=(X_1,\ldots,X_n).
\]
We wish to compress $\bo X$ into an $nR$-bit codeword
\[
\bo C=(C_1,\ldots,C_{nR})
\]
such that each $X_i$ can be reliably recovered from a
subset of the codeword bits, while those bits reveal no
information about $\bo X$ beyond what $X_i$ itself reveals.

In prior work~\cite{sid2025simple,chandar2026compression},
we showed that memoryless sources can be compressed at
any rate above entropy while satisfying these local
privacy and reliability requirements. The techniques
developed there are specific to memoryless sources.  Here, we show that every stationary ergodic
source can be compressed at any rate above its entropy
rate while preserving private local decoding. We also show that
there exist non-ergodic sources for which private local
decoding incurs a strictly positive rate penalty.

To establish the achievability result, we construct a code based on representing the source sequence, with high probability, as the sum of a
concatenation of typical blocks and a sparse error
vector. The typical-block component admits
private local decoding at rates arbitrarily close
to the entropy rate using a randomized lookup table
shared across blocks. The error vector, in turn,
admits private local decoding at an arbitrarily
small rate through an expander-based decomposition
into block-sparse components that can be efficiently
encoded using shared lookup tables. Finally, a
randomized sumset construction allows us to combine
private encodings of the components into a private
encoding of their sum. This construction is used
both to encode the sparse error vector and to combine
it with the typical-block component.

\subsection*{Related work}

\emph{Local decoding and compressed data structures.}
Locally decodable source codes and compressed data
structures share a common goal: compact representations
that permit access to individual symbols without
decompressing the entire sequence. In information
theory, the emphasis has been on the tradeoff between
compression rate and the number of codeword bits
accessed to recover a source symbol. This tradeoff
was studied by Makhdoumi \emph{et al.}~
\cite{makhdoumi_locally_2015}, Mazumdar \emph{et al.}~
\cite{mazumdar2015local}, and Tatwawadi \emph{et al.}~
\cite{tatwawadi_universal_2018}, with extensions to
variable-length coding~\cite{pananjady2018effect},
local updates~\cite{vatedka2019local}, and distributed
compression~\cite{vatedka2023local}. Work on compressed
data structures has focused on storage space and query
efficiency, measured in computation time or memory
probes~\cite{raman2007succinct,ferragina2007storage,
bille2011random}.

\emph{Expander graphs.}
Expander graphs have found applications in
error-correcting codes, sparse recovery, and
pseudorandomness; see~\cite{hoory2006expander}
for a survey. In particular, Mazumdar \emph{et al.}~
\cite{mazumdar2015local} use expanders for local
recovery from compressed representations. Here,
we exploit expansion to decompose sparse vectors
into block-sparse components that admit
space-efficient private encoding using shared
lookup tables.

\subsection*{Paper organization}

Section~\ref{sec:introduction} introduces the privacy and reliability requirements and states the main result. Section~\ref{lapro} contains the proof, first for binary sources and then for general finite alphabets.  Section~\ref{esten} discusses extensions to the private recovery of multiple source symbols. Section~\ref{impo} shows that, without the stationary ergodic assumption, private local decoding can incur a nonvanishing compression-rate penalty. Finally, Section~\ref{concl} provides concluding remarks.
\section{Preliminaries and main result}\label{sec:introduction}

We recall the definitions of locally decodable codes and locally private codes from~\cite{chandar2024entropy}.

\begin{definition}[Locally decodable code]\label{locdeco}
Fix a finite alphabet $\cX$, a blocklength $n\geq 1$,
and a rate $R>0$. A rate-$R$ \emph{locally decodable
code} $\cC_n$ consists of:
\begin{itemize}
    \item a randomized encoder that maps each
    $\bfx\in\cX^n$ to a codeword
    $\bo C=(C_1,\ldots,C_{nR})$ according to a
    distribution $P_{\bo C|\bfx}$, where
    $C_j\in\{0,1\}$;%
    \footnote{We write $nR$ for $\lceil nR\rceil$.}

    \item for each $i\in[n]$, an index set
    $\cI_i\subseteq[nR]$ and a deterministic
    decoder
    \[
    f_i:\{0,1\}^{|\cI_i|}\to\cX
    \]
    that produces the estimate
    \[
    \hat X_i=f_i(\bo C_{\cI_i}),
    \]
    where $\bo C_{\cI_i}$ denotes the
    codeword components indexed by $\cI_i$.
\end{itemize}
Given a source $\{X_i\}_{i\geq 1}$ over $\cX$,
let $\bo X=(X_1,\ldots,X_n)$ and
$\hat{\bo X}=(\hat X_1,\ldots,\hat X_n)$.
The error probability of $\cC_n$ is
\[
P_e^{(n)}=\Pr(\hat{\bo X}\ne\bo X).
\]
\end{definition}

\begin{definition}[Reliability and privacy]\label{defrelpri}
Given a source $\{X_i\}_{i\ge 1}$ over $\cX$, rate $R$ is said to be achievable with private local decoding if there
exist rate-$R$ locally decodable codes $\{\mathcal C_n\}$
and sets $\{\cS_n\}$, with $\cS_n\subseteq\cX^n$, such that
\[
P_e^{(n)}=o(1),
\qquad
\Pr(\bo X\in\cS_n)=1-o(1),
\]
and, for every $i\in[n]$ and every
$\bfx,\bfx'\in\cS_n$ with $x_i=x_i'$,
\begin{equation*}
P_{\bo C_{\cI_i}\mid\bo X=\bfx}
=
P_{\bo C_{\cI_i}\mid\bo X=\bfx'}.
\end{equation*}
Equivalently, for each $i\in[n]$, there exist distributions
$Q_{i,a}$, $a\in\cX$, such that
\[
P_{\bo C_{\cI_i}\mid\bo X=\bfx}=Q_{i,x_i},
\qquad \bfx\in\cS_n.
\]
\end{definition}

The privacy condition means that, on $\cS_n$, the
distribution of the bits accessed to recover $X_i$
depends only on $X_i$. Thus, these bits cannot
distinguish between sequences in $\cS_n$ that agree
on coordinate $i$. The restriction to a high-probability
set is essential: requiring privacy for every source
sequence would in general preclude nontrivial compression,
as shown in Appendix~B.

For our construction, it will be convenient to consider
sets on which both privacy and error-free decoding hold.

\begin{definition}[Zero-error privately decodable set]
A set $\mathcal R\subseteq\mathcal X^n$ is said to be
\emph{zero-error privately decodable at rate $R$}
if there exists a rate-$R$ locally decodable code
such that, for every $\bo x\in\mathcal R$,
\[
\Pr(\hat{\bo X}=\bo x\mid\bo X=\bo x)=1,
\]
and, for every $i\in[n]$ and
$\bfx,\bfx'\in\mathcal R$ with $x_i=x_i'$,
\[
P_{\bo C_{\cI_i}\mid\bo X=\bfx}
=
P_{\bo C_{\cI_i}\mid\bo X=\bfx'}.
\]
\end{definition}

\begin{remark}
Private local decoding can be viewed as a packing
problem: the local observations must be accommodated
within $nR$ codeword bits while satisfying privacy
and reliability. In terms of the distributions
$Q_{i,a}$ introduced in Definition~\ref{defrelpri}, privacy requires that,
for every $\bo x$ in the prescribed set~$\mathcal{S}_n$, the marginals
\[
\{Q_{i,x_i}:i\in[n]\}
\]
be jointly realizable by a distribution
$P_{\bo C\mid\bo X=\bo x}$.
Moreover, each local observation must allow recovery
of the corresponding source symbol.
\end{remark}

Our main result says that, for stationary ergodic sources, private local decoding entails no asymptotic rate penalty.

\begin{theorem}\label{main}
Let $\{X_i\}_{i\geq 1}$ be a stationary ergodic source
over a finite alphabet $\mathcal X$, with entropy rate
$\mathcal H$. For every $R>\mathcal H$, there exist a sequence of rate-$R$ zero-error privately decodable sets $\{\cS_n\}$ such that
\[
\Pr(\bo X\in\cS_n)=1-o(1).
\]
If the source is finite-order Markov, the sets $\cS_n$
can be chosen so that
\[
\Pr(\bo X\in\cS_n)\geq 1-e^{-cn}
\]
for some $c>0$ and all sufficiently large $n$.
\end{theorem}

Theorem~\ref{main} strengthens the achievability requirement of Definition~\ref{defrelpri} by guaranteeing error-free decoding on $\cS_n$. In particular,
\[
P_e^{(n)}\leq\Pr(\bo X\notin\cS_n),
\]
whereas Definition~\ref{defrelpri} allows decoding errors even on $\cS_n$. 

The stronger reliability guarantee yields a further consequence. Let $E_n=\{\bo X\in\cS_n\}$. Since $X_i$ can be recovered without error from $\bo C_{\cI_i}$ on $E_n$,
\[
I(\bo X;\bo C_{\cI_i}\mid E_n)
\geq I(X_i;\bo C_{\cI_i}\mid E_n)
=H(X_i\mid E_n).
\]
On the other hand, privacy implies
\[
I(\bo X;\bo C_{\cI_i}\mid E_n)
=I(X_i;\bo C_{\cI_i}\mid E_n)
\leq H(X_i\mid E_n).
\]
Hence, on $\cS_n$, the local observation reveals exactly the minimum amount of information about the source required for the error-free decoding of $X_i$.

Although privacy is guaranteed only on $\cS_n$, it also bounds unconditional information leakage. Let $\alpha_n\defeq\Pr(\bo X\notin\cS_n)$. Since
\[
I(\bo X_{-i};\bo C_{\cI_i}\mid X_i,\bo X\in\cS_n)=0,
\]
we have\footnote{$\log$ is to the base two.}
\[
I(\bo X_{-i};\bo C_{\cI_i}\mid X_i)
\leq h_B(\alpha_n)+\alpha_n(n-1)\log|\cX|.
\]
Thus, the normalized leakage vanishes since $\alpha_n=o(1)$. For finite-order Markov sources, Theorem~\ref{main} guarantees exponentially small $\alpha_n$, so even the unnormalized leakage vanishes.
\section{Proof of Theorem~\ref{main}}
\label{lapro}
The bulk of the proof of Theorem~\ref{main} is devoted
to binary sources, treated in Sections~\ref{lut}--\ref{dante}.
The extensions to nonbinary and Markov sources require
only minor modifications, and are presented in
Sections~\ref{nonbinar} and~\ref{expdec}, respectively.

The proof of Theorem~\ref{main} for binary sources proceeds in three steps:
\begin{itemize}
    \item In Section~\ref{lut}, we show that any set
    $\mathcal T\subseteq\{0,1\}^b$ can be made zero-error privately
    decodable using a randomized lookup table (LUT). At this stage, the
    concern is privacy rather than space complexity.

   \item In Section~\ref{amorto}, we amortize the cost of a randomized LUT by sharing it across disjoint source blocks. This allows us to privately encode concatenations of typical blocks at rates arbitrarily close to the source entropy rate. 

\item In Section~\ref{dante}, we establish a private sumset construction that preserves privacy when adding separately privately decodable components. We then show that sparse error vectors can be privately encoded at arbitrarily small rates by decomposing them, using expander graphs, into block-sparse components that admit space-efficient private representations through shared LUTs. The sumset construction allows us to add these components while preserving privacy. Finally, in the proof of Theorem~\ref{main}, we show that, with high probability, a source sequence can be expressed as the sum of a concatenation of typical blocks and a sparse error vector, and privately encode this sum at the desired rate.
\end{itemize}

\subsection{LUT}\label{lut}
Let $\mathcal T\subseteq\{0,1\}^b$, viewed as an
$|\mathcal T|\times b$ table whose rows are the elements of $\mathcal T$.
Choose a uniformly random permutation $\sigma$ of the rows and denote the
resulting table by $\widetilde{\mathcal T}$. 

The codeword assigned to $x^b\in\mathcal T$ consists of
$\widetilde{\mathcal T}$ together with the row index of $x^b$ in
$\widetilde{\mathcal T}$, say $j$. Hence, the rate is 
\begin{align}\label{noamort}
\frac{1}{b}
\bigl(b|\mathcal T|+\lceil\log|\mathcal T|\rceil \bigr)
=
\underbrace{|\mathcal T|}_{\text{privacy}}
+
\underbrace{\frac{1}{b}\lceil \log|\mathcal T|\rceil }_{\text{compression}} .
\end{align}

To recover $x_i$, decoder $i$ reads the row index $j$
and the entire $i$th column of $\widetilde{\mathcal T}$,
and outputs its $j$th entry. By symmetry of the
random permutation, the joint distribution of $j$
and this column depends on $\bfx$ only through $x_i$.
Thus, decoding is error-free and private.\footnote{
Alternatively, decoder $i$ could read $j$ and only
the $i$th entry of row $j$. However, this would
require choosing the accessed table position after
observing $j$, whereas Definition~\ref{locdeco}
requires a fixed set of codeword coordinates.
Reading the entire column avoids this adaptivity.}

\subsection{Shared-LUT}\label{amorto}
Fix $\mathcal T\subseteq\{0,1\}^b$, let $n$ be a multiple
of $b$, and consider the set $\mathcal T^{n/b}$ of
sequences composed of $n/b$ consecutive length-$b$
blocks, each in $\mathcal T$. We encode all blocks using
the same randomly permuted table $\widetilde{\mathcal T}$,
storing once the table itself and, for each block, its
row index in $\widetilde{\mathcal T}$.

To recover $x_i$, write $i=(k-1)b+r$, where
$k\in[n/b]$ and $r\in[b]$. Decoder $i$ reads the row
index of block $k$ and the entire $r$th column of
$\widetilde{\mathcal T}$, and outputs the entry at
their intersection. Decoding is error-free, and
privacy follows as in Section~\ref{lut}.

The resulting shared-LUT scheme with base set
$\mathcal T$ is zero-error and private, and operates
at rate
\begin{align}\label{amort}
\frac{1}{n}
\left(
b|\mathcal T|
+\frac{n}{b}\lceil\log|\mathcal T|\rceil
\right)
=
\underbrace{\frac{b}{n}|\mathcal T|}_{\text{privacy}}
+
\underbrace{\frac{1}{b}\lceil\log|\mathcal T|\rceil}
_{\text{compression}}.
\end{align}
The privacy cost induiced by the LUT is paid only once and
is amortized across all blocks.

 The following result is well known (see, {\it{e.g.}}, \cite[Corollary 12.7]{polyanskiy2025information}):
\begin{lemma}[Typicality]\label{ergo} Let $\{X_i\}_{i\geq 1}$ be a stationary ergodic source
over a finite alphabet $\mathcal X$, with entropy rate
$\mathcal H$.  Define the typicality set
$${\mathcal{T}}_\varepsilon^b\defeq \left\{x^b: \left|\frac{1}{b}\log \frac{1}{P_{X^b}(x^b)}-{\mathcal{H}}\right|\leq \varepsilon\right\}.$$
For any $\varepsilon>0$  we have
\begin{itemize}
    \item[i.] $ |{\mathcal{T}}_\varepsilon^b|\leq 2^{b({\mathcal{H}}+\varepsilon)}$, 
 \item[ii.] $\Pr(X^b\in {\mathcal{T}}_\varepsilon^b)\geq 1-\eta_b$
  where $\eta_b\overset{b\to \infty}{\longrightarrow} 0$. 
\end{itemize}
\end{lemma}
Equation \eqref{amort} and Lemma~\ref{ergo}.i give:
\begin{lemma}[Typicality-Shared-LUT]\label{lexe}
Let $b$ be a fixed positive integer, $n$ a multiple of $b$, and $\varepsilon\in (0,1)$. The shared-LUT scheme with base set ${\mathcal T_\varepsilon^b}$ is zero-error, private, and achieves rate
$$\frac{1}{b}\lceil \log |{\mathcal T_\varepsilon^b}|\rceil+\frac{b}{n}|{\mathcal T_\varepsilon^b}|\leq \mathcal{H}+\varepsilon+\frac{1}{b}+\frac{b2^{b({\mathcal{H}}+\varepsilon)}}{n}.$$
\end{lemma}

\subsection{Private compression of sumsets and sparse errors}\label{dante}

We first establish a construction for privately encoding sums of separately privately decodable components. Simply concatenating their encodings need not preserve privacy: the local observation may reveal the individual summands rather than only their sum. The construction below avoids this additional disclosure through random masking.

Throughout, we use ``$+$'' for ordinary addition, addition in $\mathbb Z_q$, and sumsets, with its meaning clear from context.

\begin{definition}[Closure under complementation]\label{complemento}
The closure under complementation of a set ${\mathcal T}\subseteq\{0,1\}^\ell$ is defined as
\[
\bar{\mathcal T}\defeq {\mathcal T}\cup\{\bar{\bo{t}}:\bo{t}\in{\mathcal T}\},
\]
where $\bar{\bo{t}}={\bo{t}}+\bo 1$ denotes the bitwise complement of $\bo{t}$.
The set ${\mathcal T}$ is said to be \emph{closed under complementation} if
$\bar{\mathcal T}={\mathcal T}$.
\end{definition}

\begin{proposition}[Private sumset compression via masking]\label{prop1}
Suppose $\bar{\cT},\bar{\cU}\subseteq\{0,1\}^\ell$ are zero-error privately decodable sets at rates $R_{\bar{\cT}},R_{\bar{\cU}}$ respectively. Then their sumset $$\bar{\mathcal{T}}+\bar{\mathcal{U}}\defeq\{\bo{x}:\bo{x}=\bo{t}+\bo{u},\bo{t}\in \bar{\mathcal{T}}, \bo{u}\in \bar{\mathcal{U}}\}$$ admits a zero-error locally private compressor operating at rate $R_{\bar{\mathcal{T}}}+R_{\bar{\mathcal{U}}}$.
\end{proposition}

\begin{proof}
Let $\mathrm{Enc}^{\bar{\cT}}(\cdot)$ and
$\mathrm{Enc}^{\bar{\cU}}(\cdot)$ be the encoding maps
associated with $\bar{\cT}$ and $\bar{\cU}$,
respectively. To encode
$\bo x\in\bar{\cT}+\bar{\cU}$, pick any
$\bo t\in\bar{\cT}$ and $\bo u\in\bar{\cU}$
such that $\bo t+\bo u=\bo x$, and let
$\bo V=(V_1,\ldots,V_\ell)$ be the all-zero or
all-one sequence, each with probability $1/2$,
independently of $(\bo t,\bo u)$. 
Define the combined encoder using the same random mask \(\bo V\) in both components (but independent internal randomizations for the two component encoders)
\[
\mathrm{Enc}^{\bar{\cT}+\bar{\cU}}(\bo x)
\defeq
\bigl(
\mathrm{Enc}^{\bar{\cT}}(\bo t+\bo V),
\mathrm{Enc}^{\bar{\cU}}(\bo u-\bo V)
\bigr),
\]
This is well-defined since $\bar{\cT}$ and
$\bar{\cU}$ are closed under complementation.

To decode $x_i$, apply the two local decoders to
recover $t_i+V_i$ and $u_i-V_i$, respectively,
and add them to obtain $x_i$. Thus decoding is
error-free.

To verify privacy, let $Q^{\bar{\cT}}_{i,a}$
and $Q^{\bar{\cU}}_{i,a}$ denote the distributions
of the respective local observations when the
$i$th input symbol is $a\in\{0,1\}$.
By privacy of the component codes, these
distributions do not depend on the other input
symbols.

Given $\bo X=\bo x$, the symbol $t_i+V_i$
is uniform over $\{0,1\}$, while
$u_i-V_i=x_i-(t_i+V_i)$.
Since the component encoders use independent
internal randomness, the joint distribution
of their local observations is
\[
\frac12\sum_{a\in\{0,1\}}
Q^{\bar{\cT}}_{i,a}
\otimes Q^{\bar{\cU}}_{i,x_i-a}.
\]
The expression depends on $\bo x$ only through
$x_i$, establishing privacy.
\end{proof}

\begin{corollary}[Generalization to multiple summands]
\label{multisum}
Let $\bar{\mathcal U}_1,\ldots,\bar{\mathcal U}_K\subseteq
\{0,1\}^\ell$ be zero-error privately decodable
sets at rates
$R_1,\ldots,R_K$, respectively. Then the sumset
\[
\bar{\mathcal U}_1+\cdots+\bar{\mathcal U}_K
\]
is zero-error privately decodable at rate
$\sum_{k=1}^K R_k$.
\end{corollary}

\begin{proof}
We illustrate the construction for $K=4$.
Let $\mathrm{Enc}_k$ and $f_{k,i}$ denote the
encoder and the $i$th local decoder associated
with $\bar{\mathcal U}_k$.

Given
\[
\bo e=\bo s^{(1)}+\bo s^{(2)}
+\bo s^{(3)}+\bo s^{(4)},
\qquad \bo s^{(k)}\in\bar{\mathcal U}_k,
\]
let $\bo V_1,\bo V_2,\bo V_3$ be independent,
each uniform over $\{\bo 0,\bo 1\}$.
Encode $\bo e$ as
\[
\begin{aligned}
\mathrm{Enc}(\bo e)=\bigl(&
\mathrm{Enc}_1(\bo s^{(1)}+\bo V_1+\bo V_2+\bo V_3),\\
&
\mathrm{Enc}_2(\bo s^{(2)}-\bo V_1),\\
&
\mathrm{Enc}_3(\bo s^{(3)}-\bo V_2),\\
&
\mathrm{Enc}_4(\bo s^{(4)}-\bo V_3)
\bigr).
\end{aligned}
\]
Closure under complementation ensures that all
encoder inputs belong to their respective sets.

To recover $e_i$, decoder $i$ applies $f_{k,i}$
to the corresponding local observation from
the $k$th codeword, obtaining
\[
\begin{aligned}
Y_i^{(1)}&=s_i^{(1)}+V_{1,i}+V_{2,i}+V_{3,i},\\
Y_i^{(2)}&=s_i^{(2)}-V_{1,i},\\
Y_i^{(3)}&=s_i^{(3)}-V_{2,i},\\
Y_i^{(4)}&=s_i^{(4)}-V_{3,i}.
\end{aligned}
\]
It then outputs
\[
\hat e_i=\sum_{m=1}^4Y_i^{(k)}=e_i.
\]
Thus decoding is error-free.

For general $K$, the construction uses $K-1$
independent masks $\bo V_1,\ldots,\bo V_{K-1}$,
and separately encodes
\[
\bo s^{(1)}+\sum_{j=1}^{K-1}\bo V_j,
\qquad
\bo s^{(k)}-\bo V_{k-1},
\quad 2\leq k\leq K.
\]

The same argument as for the two-summand case establishes zero-error decoding
and privacy: the $K$ local observations have joint
distribution
\[
\frac{1}{2^{K-1}}
\sum_{\substack{a_1,\ldots,a_K\in\{0,1\}\\
                 \sum_{k=1}^K a_k=e_i}}
\bigotimes_{k=1}^K Q^{\bar{\mathcal U}_k}_{i,a_k},
\]
which depends only on $e_i$. The total rate is
$\sum_{k=1}^K R_k$.
\end{proof}

The following result shows that sparse error vectors admit private encodings at arbitrarily small rates.

\begin{proposition}[Sparse sequence compression]\label{spa}
   For any fixed $0< \delta\leq 10^{-5}$, and for sufficiently large $n$, the set 
$${\mathcal{E}}_n^\delta\defeq \{\bo{e}\in \{0,1\}^n : \|\bo{e}\|_0\leq \delta n \quad \text{or}\: \geq (1-\delta) n\}$$ is zero-error privately decodable at rate 
   \begin{equation}\label{eq:R_sparse}
       R\leq C\cdot \delta\cdot (\ln (1/\delta))^3+o(1) 
   \end{equation}
   where $C>0$ is a finite constant.
\end{proposition}

We first prove Theorem~\ref{main} assuming Proposition~\ref{spa}, and then prove Proposition~\ref{spa}.
\begin{proof}[Proof of Theorem~\ref{main}]

 Suppose $\{X_i\}_{i\geq 1}$ is a stationary ergodic source
 with entropy rate
$\mathcal H$. Fix a target rate $R>\mathcal H$, and let
\[
\Delta\defeq R-\mathcal H.
\]

Let $b=b(n)$ tend to infinity sufficiently slowly
that
\[
\frac{b2^{b(\mathcal H+\varepsilon)+1}}{n}=o(1).
\]
For now, assume that $b$ divides $n$.

Define sets
\[
\mathcal T_n
\defeq
\bigl(\overline{\mathcal T_\varepsilon^b}\bigr)^{n/b}\quad\text{and}\quad
\qquad
\mathcal U_n
\defeq
\mathcal E_\delta^n,
\]
for suitable $\varepsilon,\delta>0$ to be chosen later. 
Both sets are closed under complementation. 

Since the complementation of a set at most doubles its cardinality, we have that $|\overline{\mathcal T_\varepsilon^b}|\leq 2 |{\mathcal T_\varepsilon^b} |$. Hence, by Lemma~\ref{lexe} set $\mathcal T_n$ is zero-error privately decodable at rate at most
\[
 \mathcal{H}+\varepsilon  +\frac{2}{b}+\frac{b2^{b({\mathcal{H}}+\varepsilon)+1}}{n}\leq \mathcal{H}+2\varepsilon
\]
for $n$ large enough, through the shared-LUT scheme with base set $\overline{\mathcal T_\varepsilon^b}$.

Similarly, by Proposition~\ref{spa}, set $\mathcal U_n$ is zero-error privately decodable at rate at most
\[
C\cdot \delta\cdot (\ln(1/\delta))^3+o(1).
\]
Hence, by Proposition~\ref{prop1}, the sumset
\[
\mathcal S_n
\defeq
\mathcal T_n+\mathcal U_n
\]
is zero-error privately decodable at rate at most
\[
\mathcal H+2\varepsilon
+C\cdot\delta\cdot (\ln(1/\delta))^3+o(1)
\]
as  $n\to \infty$.

By choosing $0<\delta\leq 10^{-5}$ and $\varepsilon>0$ sufficiently small that
\[
2\varepsilon+C\cdot \delta\cdot (\ln(1/\delta))^3<\Delta,
\]
we deduce that $\mathcal{S}_n$ is zero-error privately decodable at rate $\leq R$ for $n$ large enough.

Finally, we show that $\Pr( X^n\in\mathcal S_n)=1-o(1)$. Let 
$$\ell(X^n)=|\{i\in [n/b]: X_{(i-1)b+1}^{ib}\notin {\mathcal{T}}_\varepsilon^b\}|$$
denote the number of atypical blocks of $X^n$. If $\ell(X^n)\leq (n/b)\delta$, sequence $X^n$ can be written as 
$$X^n=T^n+E^n$$
where $T^n\in \mathcal{T}_n$ and $E^n\in {\mathcal{E}}_n^\delta$---since each atypical block contributes at most $b$ to the weight of $E^n$. By
Markov's inequality, Lemma~\ref{ergo}~(ii.), and stationarity  $$\Pr(\ell(X^n)\leq (n/b)\delta)\geq 1-\frac{\eta_b}{\delta}=1-o(1),$$
which proves the desired claim. 
 
Finally, the assumption that \(b\) divides \(n\) can be removed by setting \(N=b\lceil n/b\rceil\). For each length-\(n\) prefix of a sequence in \(\mathcal S_N\), fix an extension in \(\mathcal S_N\), and encode it using the length-\(N\) construction. Decoding is restricted to the first \(n\) symbols. Both privacy and zero-error decoding are preserved. Since \(N/n=1+o(1)\), the rate overhead vanishes, and the resulting rate remains below \(R\) for all sufficiently large \(n\), provided the length-\(N\) construction operates at a rate strictly below \(R\).

 This concludes the proof.
\end{proof}

\begin{proof}[Proof of Proposition~\ref{spa}]
Fix $0<\delta\leq 10^{-5}$,
and let
\[
\mathcal B_n^\delta\defeq
\{\bo e\in\{0,1\}^n:\|\bo e\|_0\leq\delta n\}.
\]
Our goal is to decompose every $\bo e\in\mathcal B_n^\delta$ into a sum
of vectors that admit low-rate private encoding by the shared-LUT scheme.
The construction will also accommodate their complements, thereby
covering the entire set $\mathcal E_\delta^n$. 

The key observation is that if a vector
can be partitioned into blocks of size $b$, each belonging to the set of sparse sequences
\[
\mathsf{Sp}_b
\defeq
\left\{\bo z\in\{0,1\}^b:\|\bo z\|_0\leq1\right\},
\]
then it can be privately encoded at small rate using shared-LUT with base set $\mathsf{Sp}_b$, since
$\mathsf{Sp}_b$ contains only $b+1$ elements (see \eqref{amort}).

The difficulty is that, although every
$\bo e\in\mathcal B_n^\delta$ has few nonzero coordinates, several of them may lie in the same block of a given partition. Can we find a small collection of partitions such that, for every $\bo e\in\mathcal B_n^\delta$, many of its nonzero coordinates are isolated in at least one partition?

Expander graphs provide a collection of partitions such that, for every sufficiently sparse vector, a constant fraction of its nonzero coordinates are the sole nonzero entry in a block of at least one partition. Peeling off these coordinates leaves a residual of smaller Hamming weight. Repeating the procedure yields a decomposition
\[
\bo e=\sum_i\bo s(i),
\]
where each $\bo s(i)$ has at most one nonzero coordinate per block of some partition. Each component admits a space-efficient private encoding using a shared LUT. By closing the corresponding sets under complementation, Proposition~\ref{prop1} allows us to combine these encodings while preserving privacy. The rate analysis shows that the total cost, including the LUT storage and the overhead due to complementation, can be made arbitrarily small.

\begin{description}
\item[Partition graph representation.]
Let $b$ be an integer and suppose $n$ is a multiple of $b$. Let  
$\mathcal P_1,\ldots,\mathcal P_d$ be $d$ partitions of $[n]$, each 
\[
\mathcal P_k
=
\{\mathcal P_{k1},\ldots,\mathcal P_{km}\}
\]
consisting of $m=n/b$ subsets of cardinality $b$. 
In the sequel we will refer to the $\mathcal P_{kj}$'s as ``blocks.''

Represent these partitions by a bipartite graph with left vertex
set $[n]$ (the variable nodes) and right vertex set $[m]$ (the
check nodes)---see Fig.~\ref{fig:expander_progress}. For each $k\in[d]$, there is an edge of color $k$ between left vertex $\ell$ and
right vertex $j$  whenever
$\ell\in\mathcal P_{k,j}$. The resulting graph has 
\begin{itemize}
    \item 
 left degree $d$ corresponding to $d$ different colored edges; 
\item degree $d\cdot b$ across check nodes;
\item possibly multiple edges between a variable node and a check node, albeit of different colors.
\end{itemize}

A block $\mathcal P_{k,j}$ is \emph{active} on a vector
$\bo r\in\{0,1\}^n$ if
$\|\bo r_{\mathcal P_{k,j}}\|_0=1$;
its unique nonzero coordinate is then said to be \emph{isolated}
by partition $\mathcal P_k$.

\begin{figure}
    \centering
    \includegraphics[width=0.3\linewidth]{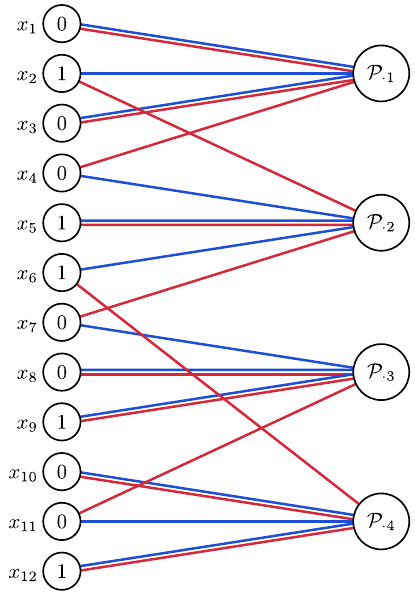}
    \caption{ Two partitions,  
$\mathcal P_{1\cdot}$ (blue) and $\mathcal P_{2\cdot}$ (red), each consisting of four blocks of size $b=3$. To illustrate the peeling procedure, suppose $\bo e=\bo r(0)=010011001001$. The blue partition isolates $x_2,x_9,x_{12}$ in blocks $\mathcal P_{1,1}$, $\mathcal P_{1,3}$, and $\mathcal P_{1,4}$, respectively. Peeling these coordinates gives
$\bo s(1)=010000001001$ and $\bo r(1)=000011000000$.
The red partition then isolates $x_5$ and $x_6$
in blocks $\mathcal P_{2,2}$ and $\mathcal P_{2,4}$,
respectively, yielding $\bo s(2)=\bo r(1)$ and $\bo r(2)=\bo 0$.
} 
    \label{fig:expander_progress}
\end{figure}
\item[One-step decomposition through peeling.]
Fix $\bo e\in\mathcal B_n^\rho$, and initialize
$\bo r(0)=\bo e$. Process the partitions successively.
At step $1\leq k \leq d$, inspect every block $\mathcal P_{k,j}$ of the
current residual $\bo r(k-1)$. If the block is active, copy it
into $\bo s(k)$ and erase it from the residual; otherwise
set the corresponding block of $\bo s(k)$ to zero and leave
the residual unchanged. Hence, if $\mathcal P_{kj}$ is active,
\[
\bo s_{\mathcal P_{kj}}(k)
=
\bo r_{\mathcal P_{kj}}(k-1),
\qquad
\bo r_{\mathcal P_{kj}}(k)=0,
\]
otherwise
\[
\bo s_{\mathcal P_{kj}}(k)=0,
\qquad
\bo r_{\mathcal P_{kj}}(k)
=
\bo r_{\mathcal P_{kj}}(k-1).
\]

Thus
\[
\bo e=\sum_{k=1}^d\bo s(k)+\bo r(d),
\]
where each $\bo s(k)$ has at most
one nonzero coordinate per block of $\mathcal P_k$.
In particular, defining the block-sparse set
\[
\mathcal D_k\defeq
\left\{\bo z\in\{0,1\}^n:
\ \bo z_{\mathcal P_{k,j}}\in\mathsf{Sp}_b
\text{ for all }j\right\},
\]
we have $\bo s(k)\in\mathcal D_k$. 
Figure~\ref{fig:expander_progress} illustrates the peeling
procedure.
\item[Guaranteed progress through expansion.]
Suppose the bipartite graph representing the partitions is an
$\left(n,m,d,\rho,\frac{3d}{4}\right)$ expander, with expansion
factor $3d/4$.\footnote{An $(n,m,d,\rho,\alpha)$ expander is a bipartite graph with $n$ left vertices with constant degree $d$, $m$ right vertices, such that every set $\mathcal A\subseteq[n]$ with $|\mathcal A|\leq\rho n$ has at least $\alpha|\mathcal A|$ distinct neighbors in $[m]$.} Let
\[
\mathcal A\defeq\supp(\bo e).
\]
Since $|\mathcal A|\leq\rho n$, the expansion property gives the lower bound on the neighbors of $\mathcal A$ 
\[
|\mathcal N(\mathcal A)|
\geq
\frac{3d}{4}|\mathcal A|.
\]
Let $\mathcal N(\mathcal A)$ denote the set of neighbors of
$\mathcal A$, and let $\mathcal N_1(A)$ denote the check nodes incident to exactly one edge from $A$---counting parallel edges separately. Since every  neighbor in $\mathcal N(\mathcal A)\backslash \mathcal N_1(\mathcal A)$ is incident to at least two edges from $A$,
\[
d|\mathcal A|
\geq
|\mathcal N_1(\mathcal A)|
+
2\bigl(|\mathcal N(\mathcal A)|-|\mathcal N_1(\mathcal A)|\bigr),
\]
and therefore
\[
|\mathcal N_1(\mathcal A)|
\geq
2|\mathcal N(\mathcal A)|-d|\mathcal A|
\geq
\frac d2|\mathcal A|.
\]
Since each left vertex has degree $d$, at least
$|\mathcal A|/2$ vertices of $\mathcal A$ are incident to a unique
neighbor. Equivalently, at least half of the nonzero coordinates of
$\bo e$ are isolated in at least one of the $d$ partitions. Each such
coordinate is peeled at some stage of the procedure above. Hence
\[
\|\bo r(d)\|_0
\leq
\frac{|\mathcal A|}{2}
\leq
\frac{\rho n}{2}.
\]
Defining
\[
\mathcal C
\defeq
\sum_{k=1}^d\mathcal D_k,
\]
we have thus shown that 
\begin{equation}\label{iter}
\mathcal B_\rho
\subseteq
\mathcal B_{\rho/2}+\mathcal C.
\end{equation}
The following lemma, whose proof is deferred to Appendix~A, gives sufficient conditions for the existence of $d$ partitions whose associated bipartite graph is an $\left(
n,m,d,\rho,\frac{3d}4
\right)$ expander graph.

\begin{lemma}[Existence of expanders]\label{lemma:expander}
For $0<\rho\leq 10^{-5}$, define
\[
d(\rho)\defeq\left\lceil 16\ln\frac{1}{\rho}\right\rceil,
\qquad
b(\rho)\defeq
\left\lfloor\frac{1}{100\rho d(\rho)}\right\rfloor.
\]
Let
\[
d=d(\rho),\qquad b=b(\rho),\qquad n=mb,\qquad m\in \mathbb{N}^+.
\]
Then there exist $d$ partitions
\[
\mathcal P_k=\{\mathcal P_{k,1},\ldots,
\mathcal P_{k,m}\},\qquad k\in[d],
\]
of $[n]$, each consisting of $m$ blocks of size $b$,
such that the associated bipartite graph
is an $(n,m,d,\rho,3d/4)$ expander.

\end{lemma}
The decomposition \eqref{iter} assumes that $n$ is a multiple
of $b$. We now extend it to arbitrary $n$ by padding,
peeling, and restricting.

Fix $0<\rho\leq\delta$, let $d=d(\rho)$ and $b=b(\rho)$,
and set
\[
N\defeq b\left\lceil\frac nb\right\rceil.
\]
By Lemma~\ref{lemma:expander}, there exist $d$ partitions of $[N]$ into blocks of size $b$ whose associated graph has expansion factor $3d/4$.

Given $\bo e\in\mathcal B_n^\rho$, append $N-n$ zeros
to obtain $\bo e^+\defeq \bo e 0^{N-n}\in\{0,1\}^N$. Since
$\|\bo e^+\|_0\leq\rho n\leq\rho N$, the peeling
procedure applies and yields
\[
\bo e^+=\sum_{k=1}^d\bo s(k)+\bo r(d),
\qquad
\|\bo r(d)\|_0\leq\frac12\|\bo e\|_0.
\]
Restricting this decomposition to the first $n$
coordinates gives
\[
\bo e=\sum_{k=1}^d\bo s(k)_{[n]}
+\bo r(d)_{[n]}.
\]
where $\bo s(k)_{[n]}$ belongs to
\[
\mathcal D_k\defeq
\left\{\bo z_{[n]}:\bo z\in\{0,1\}^N,
\ \bo z_{\mathcal P_{k,j}}\in\mathsf{Sp}_b
\text{ for all }j\right\}.
\]

As $\|\bo r(d)_{[n]}\|_0\leq\rho n/2$, we obtain
\begin{equation}\label{eq:padded-step}
\mathcal B_n^\rho
\subseteq
\mathcal B_{\rho/2}^n+
\sum_{k=1}^d\mathcal D_k.
\end{equation}
which now holds for arbitrary $n$.

\item[Iteration.]
We apply the preceding padding–peeling–restriction procedure repeatedly,  obtaining a decomposition  in \(\{0,1\}^n\) and halving the residual Hamming weight at each round. 

To keep the
total LUT storage sublinear in $n$, we use new expanders
with increasing block sizes until $b$ reaches order
$n^{1/3}$, and reuse the last expander thereafter.

Set\footnote{Recall that $0<\delta<10^{-5}$ is fixed.}
\[
\rho_i\defeq\frac\delta{2^i},\qquad
L\defeq\min\{i\geq0:n\rho_i<1\},
\]
and
\[
i_*\defeq\min\{i\geq0:b(\rho_i)\geq n^{1/3}\}.
\]
Notice that for all sufficiently large $n$, we
have $i_*<L$. For $0\leq i\leq i_*$, choose the expander of
Lemma~\ref{lemma:expander} with
\[
d_i=d(\rho_i),\qquad b_i=b(\rho_i),\qquad
N_i=b_i\left\lceil\frac n{b_i}\right\rceil.
\]
For $i_*<i<L$, use the same parameters and partitions as
at $i_*$, with $N_i=N_{i_*}$. Denote by \[
\mathcal D_k^{(i)}\defeq
\left\{\bo z_{[n]}:\bo z\in\{0,1\}^{N_i},
\ \bo z_{\mathcal P_{k,j}^{(i)}}\in\mathsf{Sp}_b
\text{ for all }j\right\}.
\] the block-sparse set at iteration $0\leq i\leq L-1$ corresponding to partition $1\leq k \leq d_i$.

For $i\leq i_*$, the chosen expander applies to any residual
of weight at most $\rho_i n$. For $i>i_*$, the expander
chosen at $i_*$ still applies, since it expands all sets of
size at most $\rho_{i_*}N_{i_*}$ and
$\rho_i n\leq\rho_{i_*}N_{i_*}$. Thus the weight of the
residual is halved at each iteration, and
\eqref{eq:padded-step} gives
\[
\mathcal B_{\rho_i}^{n}\subseteq
\mathcal B_{\rho_{i+1}}^n+
\sum_{k=1}^{d_i}\mathcal D_k^{(i)},\qquad 0\leq i<L.
\]

Since $n\rho_L<1$, $\mathcal B_{\rho_L}^n=\{0^n\}$.
Iterating, we obtain 
\begin{equation}\label{eq:prop2_decomposition}
\mathcal B_n^\delta\subseteq
\sum_{i=0}^{L-1}\sum_{k=1}^{d_i}\mathcal D_k^{(i)}.
\end{equation}

\item[Private encoding.]
For each $i,k$, define
$\widetilde{\mathcal D}_k^{(i)}$ as $\mathcal D_k^{(i)}$,
but with $\overline{\mathsf{Sp}}_{b_i}$ in place of
$\mathsf{Sp}_{b_i}$. Then
$\mathcal D_k^{(i)}\subseteq
\widetilde{\mathcal D}_k^{(i)}$, and
$\widetilde{\mathcal D}_k^{(i)}$ is closed under
complementation. 

  Encode each $\widetilde{\mathcal D}_k^{(i)}$ set with a separate shared-LUT scheme relative to base set $\overline{\mathsf{Sp}}_{b_i}$ (one independently randomized table for each pair $(i,k)$). Applying Corollary~\ref{multisum}, we combine
these encodings into a zero-error private code for
\[
\mathcal S\defeq
\sum_{i=0}^{L-1}\sum_{k=1}^{d_i}
\widetilde{\mathcal D}_k^{(i)}.
\]
By \eqref{eq:prop2_decomposition},
$\mathcal B_n^\delta\subseteq\mathcal S$.
The corresponding masking and encoding operations
are illustrated in Fig.~\ref{fig:scheme}(b).
\item[Space complexity.]
At iteration $i$, each partition has
$m_i=\lceil n/b_i\rceil$ blocks of size $b_i$.
A shared LUT requires $2b_i(b_i+1)$ bits, and
each row index 
$\lceil\log_2(2(b_i+1))\rceil$ bits. The total
number of bits is therefore
\begin{equation}\label{eq:prop2_space}
\underbrace{\sum_{i=0}^{L-1}d_i\left[
2b_i(b_i+1)\right]}_{\text{LUTs (privacy)}}+\underbrace{\sum_{i=0}^{L-1}d_i\left[ \left\lceil\frac n{b_i}\right\rceil
\left\lceil\log_2(2(b_i+1))\right\rceil
\right]}_{\text{row-indices (compression)}}.
\end{equation}

We first bound the LUT storage cost. By the definition of $i_*$,
$b_i<n^{1/3}$ for $i<i_*$, while
$b_{i_*}=\Theta(n^{1/3})$. Indeed,
$b_{i_*-1}<n^{1/3}$ and
$\rho_{i_*}=\rho_{i_*-1}/2$ imply
\[
b_{i_*}\leq2b_{i_*-1}+1=O(n^{1/3}).
\]
Since the parameters are fixed after $i_*$,
$b_i=O(n^{1/3})$ throughout. Also,
$L=O(\log n)$ and $d_i=O(\log n)$. Thus the LUT
contribution to \eqref{eq:prop2_space} is
\[
O(n^{2/3}\log^2n)=o(n).
\]

Next, using $\lceil n/b_i\rceil\leq n/b_i+1$,
the normalized row-index cost is at most
\[
\sum_{i=0}^{L-1}\frac{d_i}{b_i}
\left\lceil\log_2(2(b_i+1))\right\rceil+o(1),
\]
where the $o(1)$ term is $O(\log^3n/n)$.
For $i<i_*$, Lemma~\ref{lemma:expander} gives
\[
\frac1{b_i}\leq200\rho_i d_i,\qquad
d_i=O\left(\ln\frac1{\rho_i}\right),
\]
so
\[
\frac{d_i}{b_i}
\left\lceil\log_2(2(b_i+1))\right\rceil
=O\left(\rho_i\ln^3\frac1{\rho_i}\right).
\]
Since $\rho_i=\delta2^{-i}$,
\begin{align*}
\sum_{i<i_*}\rho_i\ln^3\frac1{\rho_i}
&\leq\delta\sum_{i\geq0}2^{-i}
\left(\ln\frac1\delta+i\ln2\right)^3\\
&=O\left(\delta\ln^3\frac1\delta\right).
\end{align*}
For $i_*\leq i<L$, the parameters are fixed at
$b_{i_*}=\Theta(n^{1/3})$ and $d_{i_*}=O(\log n)$.
The remaining contribution is therefore
\[
O\left(\frac{\log^3n}{n^{1/3}}\right)=o(1).
\]
Combining the LUT and row-index costs, the rate is at most
\[
C\delta\ln^3\left(\frac1\delta\right)+o(1)
\]
for a universal constant $C$.

\item[High-weight vectors.]
Since every $\widetilde{\mathcal D}_k^{(i)}$ is closed under
complementation, so is their sumset $\mathcal S$.
Since $\mathcal B_n^\delta\subseteq\mathcal S$, we also have
$\overline{\mathcal B}_\delta^n\subseteq\mathcal S$.
Since $\overline{\mathcal B}_\delta^n=\mathcal E_\delta^n$ we deduce that $\mathcal E_\delta^n$ is zero-error privately
decodable at rate
$C\delta\ln^3(1/\delta)+o(1)$.
\end{description}
\end{proof}

Figure~\ref{fig:scheme} summarizes the encoding
and local decoding procedures used to establish
Theorem~\ref{main}. Panel (a) shows how the
typical-block and sparse-error components are
combined, while panel (b) details the construction
of the sparse-error encoder.
\begin{figure}
    \centering
    \includegraphics[width=1\linewidth]{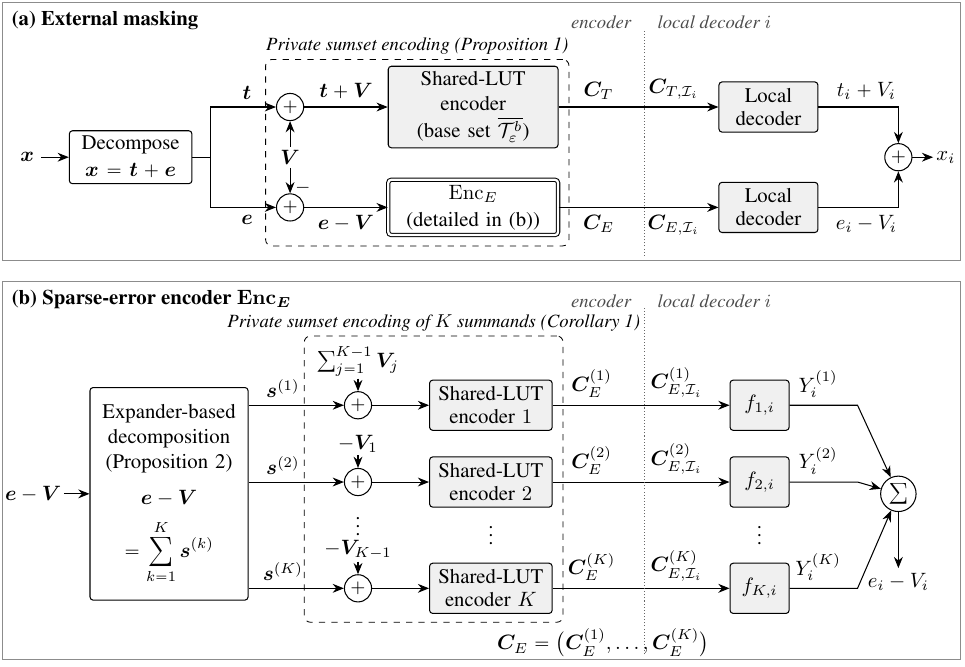}
   \caption{Overview of the private compression scheme
for binary sources.
(a) External masking: a source sequence is decomposed
as $\bo x=\bo t+\bo e$, where $\bo t$ is a
concatenation of typical blocks and $\bo e$ is sparse---if such a decomposition does not exist, declare an error.
A common random mask $\bo V$ is used to encode
$\bo t+\bo V$ and $\bo e-\bo V$ separately.
The corresponding local decoders recover $t_i+V_i$
and $e_i-V_i$, whose sum equals $x_i$.
(b) Internal construction of the sparse-error encoder
$\mathrm{Enc}_E$, highlighted in (a).
The expander-based construction decomposes the error
into $K$ components. Using $K-1$ independent random
masks, these components are privately encoded through
separate shared lookup tables.
Their local decoders recover symbols whose sum
equals $e_i-V_i$.}
\label{fig:scheme}
\end{figure}

\subsection{Nonbinary sources}
\label{nonbinar}
The extension to sources over a finite alphabet \(\mathcal X\) follows the same steps as in the binary case. We identify \(\mathcal X\) with the additive group \(\mathbb Z_q\), where \(q=|\mathcal X|\), and highlight only the differences.

Closure under complementation is replaced by ``closure under constant shifts,'' that is, for \({\mathcal{T}}\subseteq{\mathcal{X}}^b\)
\[
\overline {\mathcal{T}}=\{t+s\mathbf 1^b:t\in \mathcal{T},\ s\in\mathcal{X}\}.
\]
Proposition~\ref{prop1} then extends by replacing the binary random vector \(V\) with \(V=S\mathbf 1^\ell\), where \(S\) is uniform over \(\mathcal X\). Given a decomposition \(x=t+u\), the encoder independently encodes \(t+V\) and \(u-V\), whose sum equals \(x\).

Proposition~\ref{spa} extends by redefining the single error set $\mathsf{Sp}_b$ as  
\[
\mathsf{Sp}_b
\defeq
\left\{\bo z\in{\mathcal{X}}^b:\|\bo z\|_0\leq1\right\},
\]
and using its closure under constant shifts in the shared-LUT construction. Since
\begin{align*}
|\overline{\mathsf{Sp}}_b|
&= \left|\bigcup_{s\in\mathcal X}
\left(\mathsf{Sp}_b+s\mathbf{1}^b\right)\right|\\
&\leq \sum_{s\in\mathcal X}|\mathsf{Sp}_b|\\
&= q\bigl(1+b(q-1)\bigr),
\end{align*}
the same calculation as for the binary case shows that the rate bound remains of order
\[
O\bigl(\delta\ln^3(1/\delta)\bigr)+o(1),
\]
where the implicit constant may depend on \(q\). The expander construction and peeling argument are unchanged, as they depend only on the support of the error vector \(E^n=X^n-T^n\), where subtraction is performed component-wise modulo~\(q\).

These modifications establish Theorem 1 for stationary ergodic sources over arbitrary finite alphabets.

\subsection{Markov sources}  \label{expdec}

Suppose the source is a stationary ergodic Markov chain of order $k$. Choose $b>k$ sufficiently large that Lemma~\ref{ergo} applies.
Let $Y_i$ denote the $i$th length-$b$ block,
and let $Z_i$ indicate whether $Y_i$ is atypical.
Then $Y_1,\ldots,Y_{n/b}$ is a first-order Markov chain on
a finite alphabet, and
\[
\mathbb E[Z_i]\leq\eta.
\]
Since each atypical block contributes at most
$b$ nonzero coordinates to the residual,
\[
\|E^n\|_0\leq b\sum_{i=1}^{n/b}Z_i.
\]

The following bound then follows for every $\delta>\eta$ (using, e.g.,~\cite[Theorem 3.4]{paulin2015concentration})
\[
\Pr[\Vert E^n\Vert_0\geq \delta n ] \leq 2^{-cn},
\]
for some $c>0$ that may depend on $\delta,\eta$, and the source distribution. Choosing a small enough $\eta$, using Proposition~\ref{spa} and following the same approach as for ergodic sources gives us the desired result. 
\begin{remark}

A similar two-stage scheme for i.i.d.\ Bernoulli sources was
proposed in~\cite{chandar2026compression}. As above, the first
stage compresses $X^n$ with a small distortion, and the second
compresses the resulting error $E^n$. The difference lies
in the design of the first stage: for every $i$, the conditional
distribution of $E_i$ given $X^n=x^n$ was designed to be the same for all $x^n$; therefore $E_i$ does not reveal whether $X^n$ is typical or not.

\end{remark}

\section{Private recovery of subsets of source symbols}
\label{esten}

We consider private local recovery of subsets of source
symbols from a length-$N$ sequence
$\bo X=(X_1,\ldots,X_N)$ generated by a stationary
ergodic finite-order Markov source with entropy rate
$\mathcal H$.

Assume for simplicity that $n$ divides $N$.
Partition $\bo X$ into $N/n$ consecutive blocks of length
$n$, and encode each block independently at rate
$R>\mathcal H$ using the scheme of Theorem~\ref{main}.
Since the source is finite-order Markov, the decoding
error probability of each block decays exponentially in
$n$. Thus, choosing $n=c\log N$, with $c$ a large enough constant,
ensures that all blocks are recovered correctly with
probability tending to one.

Consider a subset of source symbols
\[
X_{\cI}=(X_i:i\in\cI).
\]
If $\cI$ contains at most one coordinate from each block,
\ie
\[
|\cI\cap\{kn+1,\ldots,(k+1)n\}|\leq 1,
\qquad 0\leq k<N/n,
\]
then $X_{\cI}$ can be privately recovered by accessing the
corresponding local codeword observations. Since the blocks are
encoded independently, the joint distribution of these
observations, conditioned on $\bo X$, depends only on
$X_{\cI}$ whenever the respective blockwise privacy
conditions hold.

For $\ell\leq N/n$, the fraction of subsets
$\cI\subseteq[N]$ of size $\ell$ satisfying this
condition is
\begin{align*}
\frac{\binom{N/n}{\ell}n^\ell}{\binom{N}{\ell}}
&=\prod_{j=0}^{\ell-1}\frac{N-nj}{N-j}\\
&=\prod_{j=0}^{\ell-1}
  \left(1-\frac{(n-1)j}{N-j}\right)\\
&\geq
\left(1-\frac{n\ell}{N-\ell}\right)^\ell\\
&\geq
1-\frac{n\ell^2}{N-\ell}\\
&=1-o(1),
\end{align*}
provided that
\[
(n-1)\ell^2=o(N).
\]
In particular, this holds whenever
\[
\ell=o\left(\sqrt{\frac{N}{\log N}}\right).
\]

Thus, for $n=c\log N$ with $c$ sufficiently large, all but a vanishing fraction of subsets of size $\ell=o(\sqrt{N/\log N})$ can be privately decoded.

\section{When Private Local Decoding Incurs a Rate Penalty}
\label{impo}
Theorem~\ref{main} shows that stationary ergodic sources can be privately compressed at any rate above entropy. The same conclusion does not hold for arbitrary sources. In fact, there are sources for which private local decoding requires a rate close to twice the entropy rate (subject to the alphabet-size constraint).

\begin{lemma}\label{nonergodic}
Let $\mathcal X$ be a finite alphabet of size $q\geq 2$, and fix
\[
0<s<R<\min\{2s,\log_2q\}.
\]
There exists a sequence of source distributions on $\mathcal X^n$ with entropy $ns+o(n)$ for which rate $R$ is not achievable with private local decoding.
\end{lemma}

\begin{proof}
Fix a rate
\[
s< R<\min\{2s,\log_2 q\}.
\]
Rates below $s$ are impossible by the source-coding converse.

Let
\[
M=\lfloor 2^{ns}\rfloor,
\qquad L=\lceil nR\rceil,
\]
and consider sources uniform over subsets $\mathcal A\subseteq\mathcal X^n$ of size $M$. Every such source has entropy $\log_2M=ns+o(1)$.

The proof has two steps. We first show that privacy limits the number of codeword bits available to each decoder. We then show that there are too few such decoders to recover most sequences in every possible support $\mathcal A$.

\begin{description}
\item[Step 1: privacy limits the number of bits accessed.]
Consider a rate-$R$ code for a source uniform on a subset $\mathcal A\subseteq\mathcal X^n$ of size $M$, and write
\[
\epsilon_n\defeq\Pr(\hat{\bo X}\ne\bo X),
\qquad
\alpha_n\defeq\Pr(\bo X\notin\cS_n).
\]
By Fano's inequality,
\[
H(\bo X\mid\bo C)
\leq h_B(\epsilon_n)+\epsilon_n\log_2M.
\]
Moreover, privacy on $\cS_n$ implies
\[
I(\bo X_{-i};\bo C_{\cI_i}\mid X_i)
\leq h_B(\alpha_n)+\alpha_n(n-1)\log_2q.
\]

For every $i\in[n]$, we have
\begin{align*}
L-|\cI_i|
&\geq H(\bo C_{\cI_i^c})\\
&\geq I(\bo X_{-i};\bo C_{\cI_i^c}
       \mid X_i,\bo C_{\cI_i})\\
&=H(\bo X_{-i}\mid X_i,\bo C_{\cI_i})
  -H(\bo X_{-i}\mid X_i,\bo C)\\
&\geq H(\bo X)-H(X_i)
  -I(\bo X_{-i};\bo C_{\cI_i}\mid X_i)
  -H(\bo X\mid\bo C).
\end{align*}
Consequently,
\begin{align}
|\cI_i|
\leq{}& L-\log_2M+\log_2q \notag\\
&+h_B(\epsilon_n)+\epsilon_n\log_2M \notag\\
&+h_B(\alpha_n)+\alpha_n(n-1)\log_2q.
\label{probe_bound}
\end{align}

Choose $\gamma>0$ such that
\[
R-s+\gamma<s,
\]
and then choose $\tau\in(0,1/4)$ sufficiently small that
\[
\tau(s+\log_2q)<\gamma.
\]
Set
\[
d\defeq\lfloor n(R-s+\gamma)\rfloor.
\]
Since $L=nR+O(1)$ and $\log_2M=ns+o(1)$, \eqref{probe_bound} implies that, for all sufficiently large $n$, every rate-$R$ code satisfying
\[
\epsilon_n\leq\tau,\qquad \alpha_n\leq\tau
\]
must have
\[
|\cI_i|\leq d,\qquad i\in[n].
\]

\item[Step 2: counting local decoders.]
A collection of local decoders consists of the sets $\cI_i$ and decoding functions
\[
f_i:\{0,1\}^{|\cI_i|}\longrightarrow\mathcal X,
\qquad i\in[n].
\]
The number $N_n$ of such collections satisfying $|\cI_i|\leq d$ is bounded by
\begin{align*}
N_n
&\leq
\left(
\sum_{j=0}^d\binom Lj q^{2^j}
\right)^n\\
&\leq 2^{nL}q^{n2^d}.
\end{align*}

Each collection defines a deterministic mapping
\[
f:\{0,1\}^L\longrightarrow\mathcal X^n
\]
whose image contains at most $2^L$ sequences.
\end{description}

Now choose $\mathcal A$ uniformly among all subsets of $\mathcal X^n$ of size $M$. Fix $f$, let $\mathcal B=\operatorname{Im}(f)$ where $|\mathcal B|\leq 2^L$, and let $t=\lceil M/2\rceil$. For any $t$ distinct elements drawn from $\mathcal A$, the probability that they all belong to $\mathcal B$ is at most $(2^L/q^n)^t$. Since there are $\binom Mt$ ways to select these elements, the union bound gives
\[
\begin{aligned}
\Pr\left(|\mathcal A\cap\operatorname{Im}(f)|\geq M/2\right)
&\leq \binom Mt\left(\frac{2^L}{q^n}\right)^t\\
&\leq 2^M\left(\frac{2^L}{q^n}\right)^{M/2}.
\end{aligned}
\]
Taking a union bound over all $N_n$ local decoders,
\begin{align*}
&\Pr\left(
\exists f:\,
|\mathcal A\cap\operatorname{Im}(f)|\geq M/2
\right)\\
&\qquad\leq
2^{nL}q^{n2^d}\,
2^M\left(\frac{2^L}{q^n}\right)^{M/2}.
\end{align*}
For all sufficiently large $n$, the right-hand side is strictly less than one. Indeed, $2^d=o(M)$ since $R-s+\gamma<s$, while
\[
q^n/2^L=2^{n(\log_2q-R)-O(1)}
\]
grows exponentially since $R<\log_2q$. 

Thus, there exists a support $\mathcal A$ of size $M$ such that
\[
|\mathcal A\cap\operatorname{Im}(f)|<M/2
\]
for every collection of local decoders accessing at most $d$ codeword bits per coordinate.

Consider the source uniform on such an $\mathcal A$. If a code satisfied $\epsilon_n\leq\tau$ and $\alpha_n\leq\tau$, Step 1 would imply $|\cI_i|\leq d$ for every $i$. But then Step 2 implies that fewer than half the sequences in
$\mathcal A$ belong to $\operatorname{Im}(f)$. Since the decoder
can only output sequences in $\operatorname{Im}(f)$, we have
\[
\epsilon_n
\geq 1-\frac{|\mathcal A\cap\operatorname{Im}(f)|}{M}
>\frac12,
\]
contradicting $\epsilon_n\leq\tau<1/4$.

Thus, for every sufficiently large $n$, there exists a set $\mathcal A_n\subseteq\mathcal X^n$ of cardinality $\lfloor 2^{ns}\rfloor$ such that no rate-$R$ code for the source uniform on $\mathcal A_n$ can satisfy both the reliability and privacy requirements. These sources have entropy $ns+o(n)$, yet rate $R>s$ is not achievable with private local decoding. This proves the lemma.
\end{proof}

\section{Concluding Remarks}
\label{concl}
We have shown that every stationary ergodic source over a finite alphabet can be compressed at any rate above its entropy rate while allowing private local decoding. Moreover, decoding is error-free and privacy is exact on a set of source sequences whose probability tends to one. For finite-order Markov sources, the probability of its complement decays exponentially. Thus, for stationary ergodic sources, local privacy imposes no asymptotic penalty on compression rate. This contrasts with the non-ergodic examples of Section~V, for which such a penalty may be unavoidable.

Several questions remain open. The present result is asymptotic, and our expander construction relies on an existence argument. An important next step is to develop explicit, efficient constructions and determine their performance at finite blocklengths. In particular, it would be useful to understand the tradeoffs among compression rate, decoding complexity, the number of codeword bits accessed, and the storage overhead of the lookup tables.

Another direction is the private recovery of multiple source symbols. Privacy for individual symbols does not generally imply privacy when several local observations are combined. Section~IV gives partial results by encoding source blocks independently, but a more general question remains: can arbitrary subsets of source symbols, or contiguous substrings, be recovered while revealing no information beyond the requested symbols, without sacrificing entropy-rate compression?

\bibliographystyle{IEEEtran}
\bibliography{codes_with_locality}

\section*{Appendix A: Proof of Lemma~\ref{lemma:expander}}

Choose $\mathcal P_1,\ldots,\mathcal P_d$ independently and uniformly
among partitions of $[n]$ into $m$ indexed blocks of size $b$.
The degree claims hold by construction. We show that the expansion
condition holds with positive probability.

Fix $A\subseteq[n]$ and $Q\subseteq[m]$, with
$|A|=a$ and $|Q|=s$. For each $k\in[d]$, the union
$\bigcup_{j\in Q}\mathcal P_{k,j}$ is a uniformly random
subset of $[n]$ of size $sb$. Thus
\[
\Pr\left(A\subseteq\bigcup_{j\in Q}\mathcal P_{k,j}\right)
=\frac{\binom{sb}{a}}{\binom na}.
\]
Now $\mathcal N(A)\subseteq Q$ if and only if
$A\subseteq\bigcup_{j\in Q}\mathcal P_{k,j}$
for every $k\in[d]$. Since the partitions are
chosen independently,
\[
\Pr\bigl(\mathcal N(A)\subseteq Q\bigr)
=\left(\frac{\binom{sb}{a}}{\binom na}\right)^d
\leq\left(\frac{sb}{n}\right)^{da}.
\]

A union bound over $A$ and $Q$ now gives
\begin{align*}
&\Pr\left(\exists A\subseteq[n]:|A|=a,\;
 |\mathcal N(A)|<\frac{3d}{4}|A|\right)\\
&\quad\leq\binom n a\binom ms\left(\frac{s}{m}\right)^{da}\\
&\quad\leq\left(\frac{en}{a}\right)^a
 e^s\left(\frac{s}{m}\right)^{da-s}\\
&\quad\leq\left[\frac{en}{a}
 \left(\frac{3e^3dba}{4n}\right)^{d/4}\right]^a\\
&\quad\leq\left[\frac e\rho
 \left(\frac{3e^3}{400}\right)^{d/4}\right]^a.
\end{align*}
Here the third inequality uses $s\leq 3da/4$ and $s/m\leq
3dba/(4n)$; the last uses $a/n\leq\rho$, $d\geq4$, and
$\rho db\leq1/100$.

Write
\[
\alpha\defeq\frac e\rho\left(\frac{3e^3}{400}\right)^{d/4}.
\]
Since $3e^3/400<e^{-1}$ and $d\geq16\ln(1/\rho)$,
\[
\alpha\leq\frac e\rho e^{-d/4}
\leq e\rho^3<\frac12.
\]
Summing over $a=1,\ldots,\lfloor\rho n\rfloor$, the probability
that expansion fails for some $A$ is at most
\[
\sum_{a=1}^{\lfloor\rho n\rfloor}\alpha^a
\leq\frac{\alpha}{1-\alpha}<1.
\]
Thus some choice of the partitions satisfies the claimed expansion.

\section*{Appendix B: Why privacy for all source sequences precludes compression}

Consider an i.i.d.\ Bernoulli($p$) source, $0<p<1$, and
suppose that a rate-$R$ code satisfies $P_e^{(n)}=o(1)$.
Since $\Pr(X_i=a)>0$ for both $a\in\{0,1\}$, we have
\[
\max_{i\in[n],\,a\in\{0,1\}}
\Pr(\hat X_i\neq X_i\mid X_i=a)=o(1).
\]

Suppose now that the privacy condition holds for every $\bo x\in\{0,1\}^n$. Since the distribution of the
observation available to decoder $i$, and hence that of
$\hat X_i$, depends on $\bo x$ only through~$x_i$,
\[
\Pr(\hat X_i\neq X_i\mid\bo X=\bo x)
=
\Pr(\hat X_i\neq X_i\mid X_i=x_i).
\]
It follows that
\[
\begin{aligned}
\mathbb E\bigl[\|\hat{\bo X}-\bo X\|_0
       \mid\bo X=\bo x\bigr]
&=\sum_{i=1}^n
\Pr(\hat X_i\neq X_i\mid X_i=x_i)\\
&=o(n),
\end{aligned}
\]
uniformly over $\bo x\in\{0,1\}^n$, hence regardless of the source distribution. In particular, applying the same encoder and decoders to an i.i.d.\ Bernoulli($1/2$) source yields
\[
\frac1n\mathbb E[\|\hat{\bo X}-\bo X\|_0]=o(1).
\]
The rate-distortion converse then gives
\[
R\geq 1-h_B(o(1))=1-o(1).
\]
Hence, requiring privacy for every source sequence
precludes compression below one bit per symbol, even
when the source entropy is strictly smaller than one. The above argument extends immediately to general alphabets.

\end{document}